\documentclass[10pt]{iopart}

\usepackage{graphicx}
\usepackage{dcolumn}
\usepackage{bm}
\usepackage{iopams}
\usepackage{setstack}
\usepackage{eufrak}

\def\Xint#1{\mathchoice
   {\XXint\displaystyle\textstyle{#1}}%
   {\XXint\textstyle\scriptstyle{#1}}%
   {\XXint\scriptstyle\scriptscriptstyle{#1}}%
   {\XXint\scriptscriptstyle\scriptscriptstyle{#1}}%
   \!\int}
\def\XXint#1#2#3{{\setbox0=\hbox{$#1{#2#3}{\int}$}
     \vcenter{\hbox{$#2#3$}}\kern-.5\wd0}}

\def\dashint{\Xint-}

\begin{document}

\title[Parametric Spectral Statistics]{Parametric Spectral Statistics
in Unitary Random Matrix Ensembles: From Distribution Functions to
Intra-Level Correlations}

\author{I~E~Smolyarenko and B~D~Simons} 

\address{Cavendish Laboratory, University of Cambridge, Madingley Rd.,
Cambridge CB3 0HE, UK}

\begin{abstract} 
We establish a general framework to explore parametric statistics of
individual energy levels in unitary random matrix ensembles. For a
generic confinement potential $W(H)$, we (i) find the joint
distribution functions of the eigenvalues of $H$ and $H'=H+V$ for an
arbitrary fixed $V$ both for finite matrix size $N$ and in the
``thermodynamic'' $N\to\infty$ limit; (ii) derive many-point
parametric correlation functions of the two sets of eigenvalues and
show that they are naturally parametrised by the eigenvalues of the
{\em reactance matrix} for scattering off the ``potential'' $V$; (iii)
prove the universality of the correlation functions in unitary
ensembles with non-Gaussian non-invariant confinement potential
$W(H-V)$; (iv) establish a general scheme for exact calculation of {\em
  level-number-dependent} parametric correlation functions and apply
the scheme to the calculation of {\em intra-level} velocity
autocorrelation function and the distribution of parametric level shifts.
\end{abstract}


\eads{\mailto{is220@phy.cam.ac.uk},\mailto{bds10@cus.cam.ac.uk}}


\section{Introduction}

Parameter-dependent evolution of spectra in random matrix ensembles
has been a subject of investigation almost from the inception of the
Random Matrix Theory (RMT)~\cite{Dyson62,mehta}. Dyson~\cite{Dyson62} showed
that the joint probability distribution function of the eigenvalues in
the classic random matrix ensembles can be alternatively obtained as
an equilibrium distribution of a one-dimensional gas of classical
charges which undergo Browninan motion in fictitious time while
simultaneously logarithmically repelling each other. The same dynamics
of eigenvalues arises if the corresponding random matrices themselves
execute Browninan motion~\cite{Beenakker} according to a sequence of
mappings $H\mapsto H+dV$ where $H$ is drawn form a random matrix
ensemble and $dV$ is the elementary step of the random walk {\em in
the space of matrices} of the same symmetry as $H$.  

However, despite the theoretical interest in these types of models,
the experimental motivation in the early days of RMT was lacking since
the nuclear spectra for which RMT was originally devised \cite{porter}
are essentially immutable. Later, the range of RMT applications has
broadened to include systems such as mesoscopic quantum structures
(e.g., quantum dots) which can be subject to a variety of external
perturbations \cite{reviews}. In fact, controlled external
perturbations such as magnetic field or gate potential often serve as
important experimental tools to study statistical properties of such
systems. Thus, instead of a sequence of random steps $dV$ one is lead
to consider finite mappings $H\mapsto H'=H+V$, where $V$ is a fixed
matrix with the same symmetry as the ensemble from which the matrices
$H$ are drawn~\footnote[1]{A separate subject which is beyond the
scope of this paper concerns mappings such that the symmetry of $V$ is
lower than the symmetry of $H$.}\cite{haake,SA,efetov,list}.

While magnetic field clearly belongs to the class of extended
perturbations, in many applications the assumption that the rank of
$V$ scales with $N$ may not prove to be adequate in the RMT modelling
of mesoscopic devices. External perturbations such as an STM tip, a
change in the potential on a finger-shaped gate, or a jump in the
position of a bi- or multi-stable defect represent potentials which
are localised in space. In the RMT language localised potentials are
modelled by matrices $V$ of finite rank $r$. Such matrices can be
represented using a finite set of orthonormal complex vectors
$\left\{{\mathbf a}_\kappa\right\}_{\kappa=1}^r$ as
$V=N\sum_{\kappa=1}^rv_{\kappa}{\mathbf a}_{\kappa}\otimes {\mathbf
a}_{\kappa}^{\dagger}$. [For technical reasons related to the use of
the Itzykson -- Zuber -- Harish-Chandra (IZHC) integral~\cite{IZ}, our
consideration here is restricted to the unitary ensemble, hence the
requirement that ${\mathbf a}_{\kappa}$ are complex.]

In the standard RMT one can distinguish two classes of correlation
functions between the levels \cite{mehta}. In the first class the
levels are ``labelled'' by their positions (``energies'' in the
Hamiltonian interpretation of the random matrices), so that the
fundamental object is the many-point correlation function of the
density of states (DoS)
$\nu(E)=\sum_{\alpha=1}^N\delta\left(E-\epsilon_\alpha\right)$,
\begin{equation}
\tilde{R}_n\left(\{E\}_{i=1}^n\right)=\left\langle\prod_{i=1}^n\nu
\left(E_i\right)\right\rangle, 
\end{equation}
where the angular brackets denote averaging over the matrix ensemble.
In the bulk region of the spectrum the correlation functions are
translationally invariant with respect to the simultaneous shifts of
all energies, so that $\tilde{R}_n$ is actually a function of $n-1$ energy
differences. Such correlation functions are known to have
determinantal structures in all three classic Wigner-Dyson
ensembles. E.g., in the unitary ensemble, the $N\to\infty$ limit
results in
\begin{equation}
\label{rn}
\tilde{R}_n\left(\{E\}_{i=1}^n\right)=\det_{ij}\tilde{k}\left(\varepsilon_i-\varepsilon_j\right)
\end{equation}
where $\tilde{k}\left(\varepsilon_i-\varepsilon_j\right)=k\left(\varepsilon_i-\varepsilon_j\right)
-\delta\left(\varepsilon_i-\varepsilon_j\right)\theta\left(i-j\right)$,
$k\left(u\right)=\sin\pi u/\pi u$ is the celebrated sine kernel of the
Wigner-Dyson RMT, the dimensionless energies $\varepsilon\equiv
E/\bar{\Delta}$ are normalised by the mean
level spacing
$\bar{\Delta}=\left\langle\nu\left(E\right)\right\rangle^{-1}
\equiv\rho\left(E\right)^{-1}$, and the step-function $\theta$
is regularised so that $\theta(0)=0$. The term involving the
$\delta$-function accounts for the self-correlation of
levels. Analogous structures with more involved kernels arise in other
ensembles. While originally derived in the context of the Gaussian Unitary
Ensemble (GUE), the correlation kernel $k$ acquires the same universal
form in the bulk scaling limit for a wide class of non-Gaussian random
matrix ensembles \cite{BZ}
\begin{equation}
{\mathcal P}\left(H\right)=C_{\mathrm U}e^{-\mathrm{Tr}W(H)}, 
\end{equation}
where $W(H)$ is an even polynomial, and $C_{\mathrm{U}}$ is the
normalisation constant.

In the second class of correlation functions the levels are labelled
by their {\em numbers}, while their energies may or may not enter as
additional variables depending on the specific correlation
function. An archetypal example is the distribution of level spacings
$\mathfrak{S}_1(\omega)$ which is a particular case of the distribution
$\mathfrak{S}_q\left(\omega\right)$ of distances between a level $\alpha$ and
another level separated from $\alpha $ by $q-1$ other levels:
\begin{equation}
\label{pq}
\mathfrak{S}_q\left(\omega\right)=\bar{\Delta}\left\langle\delta\left(\epsilon_\alpha-E\right)
\delta\left(\epsilon_{\alpha+q}-E-\omega\right)\right\rangle.
\end{equation}
$\mathfrak{S}_1$ is actually a {\em correlation} function between an
arbitrary level $\alpha$ and its {\em neighbour}
$\alpha+1$. Translational invariance in the bulk of the spectrum
ensures that this correlation function does not depend on either $E$
or $\alpha$. In contrast with (\ref{rn}) no summation over $\alpha$ is
involved, so that level-number-specific information is retained, and,
indeed, near the edge of the spectrum, the same correlation function
-- now as a function of both level numbers -- describes the
distribution of distances between, say, the second lowest and the
third lowest levels.

The two types of the correlation functions are related to each other
{\em via} infinite series. Thus, an equivalent expression for
$\mathfrak{S}_1(\omega)$ without referring to specific level numbers
is the joint probability to find two levels at positions $E$ and
$E+\omega$ and {\em all other levels} in the interval
$\mathbb{R}/[E,E+\omega]$. After straightforward algebra \cite{mehta},
$\mathfrak{S}_1(\omega)$ is written as
\begin{equation}
\label{rtop}
\!\!\mathfrak{S}_1(\omega)=-\bar{\Delta}\partial^2_\omega\sum_{n=0}^{\infty}\frac{(-1)^n}{n!}
\int_E^{E+\omega}dE_1\ldots \int_E^{E+\omega}dE_n
R_n\left(\left\{E_i\right\}_{i=1}^n\right),
\end{equation}
where $R_n\left(\{E\}_{i=1}^n\right)=\det_{ij}k\left(E_i-E_j\right)$
is the part of $\tilde{R}_n$ from which the level self-correlation
terms are excluded. Combined with the determinantal structure of
$R_n$, this formula allows one to express $\mathfrak{S}_1$ in terms of the
derivative of a certain generating function,
$\mathfrak{S}_1=-\bar{\Delta}\partial^2_\omega\det_{\left[E,E+\omega\right]}
\left(\mathbb{I}-\hat{k}\right)$, where the determinant is understood
as a Fredholm determinant on the space of functions with the support
in the interval $\left[E,E+\omega\right]$, and the kernel $\hat{k}$ is
defined through its matrix elements: $\hat{k}(u,w)=k(u-w)$. Analogous
expressions can be obtained for other $P_q$'s. The Fredholm
determinant itself -- the generating function -- is the probability to
find {\em no} levels in the interval $\left[E,E+\omega\right]$. Other
$\mathfrak{S}_q$'s are similarly related to the Fredholm determinant
expressions for the probability to find $q-1$ levels in a given
interval.

Conversely, the correlation function $R_2(E,E+\omega)$ can be
expressed as an infinite sum over $\mathfrak{S}_q$'s,
\begin{equation}
\label{ptor}
R_2(E,E+\omega)=\sum_{q=-\infty}^{\infty}\mathfrak{S}_q(\omega)
\end{equation}
which is a simple restatement of the fact that the conditional (on the
existence of a level $\alpha$ at $E$) probability to find {\em any}
level at $E+\omega$ is the sum over $q$ of the (conditional)
probabilities to find {\em the} $(\alpha+q)$th level at that position.
Operationally, however, $R_n$ is usually more readily accessible than
$\mathfrak{S}_q$, so that equation (\ref{ptor}) and its analogues are
not often used in practical calculations.

In direct analogy with the duality between the two classes of
correlation functions described above for ordinary RMT, one can
introduce their generalisations in the parametric RMT. Thus, instead
of $R_n$ we are now lead to define the parametric many-point correlation
function as
\begin{equation}
\label{rnm}
R_{nm}\left(\{E_i\}_{i=1}^n,\{E'_j\}_{j=1}^m\right)=\left\langle\prod_{i=1}^n
\nu\left(E_i\right)\prod_{j=1}^m\nu'\left(E'_j\right)\right\rangle,
\end{equation}
where $\nu'\left(E\right)=\mathrm{Tr}\delta\left(E-H'\right)$. 

Except in the $r=1$ case (see below), the ``level spacing'' in the
combined set $\{E\}\cup\{E'\}$ does not have an immediate physical
interpretation. A more meaningful analogue of equation (\ref{pq}) is
the distribution of distances between a level $\alpha$ and the
parametric ``descendant'' of the level $\alpha+q$,
\begin{equation}
P_q\left(\omega,\mathfrak{x}\right)=\bar{\Delta}
\left\langle\delta\left(\epsilon_\alpha-E\right)
\delta\left(\epsilon_{\alpha+q}(\mathfrak{x})-E-\omega\right)\right\rangle,
\end{equation}
where $\mathfrak{x}$ is the set of the parameters characterising
$V$. Due to universality \cite{SA,MSS}, $\mathfrak{x}$ can be
identified with either a single parameter $x$ measuring the overall
strength of $V$ in the case of extended (infinite rank) perturbations,
or with a set of at most $r$ parameters in the finite rank case (see
Section 3). $P_0$ is simply the distribution of shifts of a {\em
single} level under the influence of a parametric perturbation. Note
the absence of summation over $\alpha$; nevertheless, in the bulk of
the spectrum $P_q$ is not a function of $\alpha$.

A related quantity which in the past has been extensively studied in
the context of parametric correlations in chaotic and disordered
systems using either numerical or approximate analytical techniques
\cite{szafer,SA,burg,keating,zakr} is the {\em single level} velocity
correlation function $C_0\left(\mathbf{x}\right)$. Within the
formalism developed in the present paper, it is convenient to
represent it as a special case of the following correlation function:
\begin{equation}
C_q\left(\mathfrak{x}\right)=\left\langle\partial_{\mathfrak{x}}
\epsilon_\alpha\left(\mathfrak{x}=0\right)\partial_{\mathfrak{x}}
\epsilon_{\alpha+q}\left(\mathfrak{x}\right)\right\rangle,
\end{equation}
which represents the correlation of responses to the perturbation of
two levels separated by a distance $\mathfrak{x}$ in the parameter
space, and by $q-1$ levels in the level sequence. 

The objective of this paper is to present a set of interrelated
developments in the parametric RMT. In the next section we construct a
formalism to determine the joint distribution functions for the
eigenvalues of $H$ and $H'$ for perturbing matrices $V$ of arbitrary
rank $r$. In particular, we concentrate on the case when $r$ is finite
in the $N\to\infty$ limit, although the results are equally applicable
in the opposite limit where we reproduce the Gaussian transition
kernel implicit in the earlier work on the subject \cite{SA,frelkanzieper}.

In Section 3.1 we use the joint distribution functions to derive the
many-point correlation function $R_{nm}$. We show that the latter
possesses the determinantal structure at arbitrary $r$ \cite{SMS},
generalizing the result previously known only in the context of
two-matrix models \cite{SA,frelkanzieper} (i.e., with $V$ itself drawn
from a random matrix ensemble), and thus corresponding to $r=N$ in the
classification adopted here. In Section 3.2 these results are applied
to prove (on a physical level of rigour) that the universality of the
Wigner-Dyson sine kernel extends to a class of non-invariant ensembles
characterised by the distribution function
$\mathcal{P}_V\left(H\right)=C_U\exp\left\{-\mathrm{Tr}\,W\left(H-V\right)\right\}$.

The final objective of this paper is to present (Section 3.3) a
general framework for the calculation of
$P_q\left(\omega,\mathfrak{x}\right)$, $C_q\left(\mathfrak{x}\right)$,
and related {\em number-dependent} correlation functions in terms of
the corresponding family of generating functions. Owing to the
determinantal structure of $R_{nm}$ \cite{SMS,SA,frelkanzieper} (see
also Section 3.1 below), the latter possess representations in terms of
Fredholm determinants of certain parameter-dependent integral kernels.


\section{Joint Distributions of Eigenvalues}

Consider a set of pairs of matrices $H$ and $H'=H+V$, with $H$ being a
random Hermitian $N\times N$ matrix $H$ drawn from the unitary
ensemble ${\mathcal P}\left(H\right)$ and $V$ is a deterministic
matrix of rank $r$.
The joint distribution of the matrices $H$ and $H'$ is
\begin{equation}
{\mathcal P}\left(H,H'\right)={\mathcal P}\left(H\right)
\delta\left(H+V-H'\right),
\end{equation}
where the matrix delta-function is understood as the product of
ordinary scalar delta-functions -- one per each of the $N^2$
independent variables of $H$. The joint distribution function of the
combined set of eigenvalues $\left\{\epsilon_\alpha\right\}_{\alpha=1}^N$ of the
matrix $H$ and $\left\{\epsilon^{\prime}_\beta\right\}_{\beta=1}^N$ of $H'$ is
obtained from the above expression by integrating out the angular
degrees of freedom:
\begin{equation}
{\mathcal P}\left(\left\{\epsilon\right\},
\left\{\epsilon^{\prime}\right\}\right) = \int
\Delta^2\left(H\right) \Delta^2\left(H'\right) d\mu\left(U\right)
d\mu\left(U'\right) 
 {\mathcal P}\left(H,H'\right),
\end{equation}
where $d\mu\left(U\right)$ is the invariant Haar measure on the
unitary group $U(N)$, and $\Delta\left(H\right)$
is the Vandermonde determinant
\begin{equation}
\Delta\left(H\right)\equiv\Delta\left(\left\{\epsilon\right\}\right)=
\prod_{1\leq \alpha < \beta\leq N} \left(\epsilon_\alpha-\epsilon_\beta\right).
\end{equation}
With the help of a Lagrange multiplier matrix $\lambda$ (it is
straightforward to show that independent components of $\lambda$ can
be arranged into a Hermitian matrix) the expression for the joint
distribution function takes the form
\begin{eqnarray}
\fl {\mathcal P}\left(\left\{\epsilon\right\},
\left\{\epsilon^{\prime}\right\}\right) = 
C_{\mathrm U}\int\Delta^2\left(H\right)
\Delta^2\left(H'\right)\Delta^2\left(\lambda\right)
d\mu\left(U\right)d\mu\left(U'\right)
\frac{d\mu\left(U_\lambda\right)}{\left(2\pi\right)^{N^2-N}} 
\prod_{k=1}^N \frac{d\lambda_k}{2\pi} \nonumber \\
 \times  \exp\left\{i\,\mathrm{Tr}\left(\lambda H+\lambda V-\lambda
H'\right)\right\} \exp\left\{-W\left(H\right)\right\},
\end{eqnarray}
where $\lambda_k$ are the eigenvalues of $\lambda$. Integration over
the angular degrees of freedom of $H$ and $H'$ can be performed with
the help of the IZHC integral \cite{IZ}
\begin{equation}
\int d\mu\left(U\right)\exp\left\{i\mathrm{Tr}AUBU^{\dagger}\right\}=
c\frac{\det_{\alpha\beta}\left[\exp\left(ia_\alpha b_\beta\right)\right]}
{\Delta\left(A\right)\Delta\left(B\right)},
\end{equation}
where $\left\{a_\alpha\right\}_{\alpha=1}^N$, $\left\{b_\beta\right\}_{\beta=1}^N$ are the
complete sets of the eigenvalues of the matrices $A$ and $B$,
respectively, and
$c=\left(i/2\right)^{\left(N^2-N\right)/2}\prod_{j=1}^Nj!$. The
remaining integral thus takes the form
\begin{eqnarray}
\label{10}
\fl {\mathcal P}\left(\left\{\epsilon\right\},
\left\{\epsilon^{\prime}\right\}\right) = 
\frac{C_{\mathrm U}\left|c\right|^2}{\left(2\pi\right)^{N^2}}\int \Delta\left(H\right)
\Delta\left(H'\right)d\mu\left(U_{\lambda}\right)\prod_{k=1}^N
d\lambda_k\exp\left\{i\mathrm{Tr}\left(\lambda V\right)\right\}
\nonumber \\
\fl \times  
\sum_{\sigma,\sigma^{\prime}}\left(-1\right)^{\sigma+\sigma'}
\exp\left\{i\sum_{\alpha=1}^N\left(\epsilon_{\alpha}\lambda_{\sigma\left(\alpha\right)}-\epsilon^{\prime}_{\alpha}
\lambda_{\sigma'\left(\alpha\right)}\right)\right\}
\exp \left\{-\sum_{\alpha =1}^NW\left(\epsilon_\alpha\right)\right\},
\end{eqnarray}
where the sum over $\sigma$ runs over all permutations of the indices
$1,...,N$, and $\left(-1\right)^{\sigma}$ denotes the signature of the
permutation. The ordering of the energy levels in the Vandermonde
determinants is assumed to be in the increasing order.  Due to the
invariance of the measure $\mu\left(U\right)$, the integral in
(\ref{10}) does not depend on the ``angular'' degrees of freedom of
$V$. The latter can thus be thought of as ``gauge'' degrees of freedom
of (\ref{10}), and integrated over with an appropriate measure
$d\mu_{(r)}\left(\left\{{\mathbf a}_\kappa\right\}\right)$.  A
convenient choice for $d\mu_{(r)}\left(\left\{{\mathbf
a}_\kappa\right\}\right)$ is the uniform distribution of all
components of ${\mathbf a}_\kappa$ constrained only by the requirement
of orthonormality.  $U_\lambda$ rotations can now be absorbed into the
rotations of $V$, and the integral over $d\mu(U_\lambda)$ gives a
constant $\mu_N=\int d\mu\left(U_{\lambda}\right)=
\left(2\pi\right)^{\left(N^2-N\right)/2}/\prod_{j=1}^{N}j!$. Equation
(\ref{10}) can now be rewritten as
\begin{eqnarray}
\label{genericp}
{\mathcal P}\left(\left\{\epsilon\right\},
\left\{\epsilon^{\prime}\right\}\right) = 
\frac{C_\mathrm{U}\left|c\right|^2\mu_N}
{\left(2\pi\right)^{N^2-N}}e^{-\sum_{\alpha=1}^NW\left(\epsilon_\alpha\right)}\Delta\left(H\right)
\Delta\left(H'\right) \nonumber \\
\times\int d\mu_{(r)}\left(\left\{{\mathbf
a}_\kappa\right\}\right)\det_{\alpha\beta} 
\delta\left(\epsilon_\alpha-\epsilon'_\beta-V_{\alpha\alpha}\right),
\end{eqnarray}
where $V_{\alpha\alpha}$ are the diagonal elements of $V=N\sum_{\kappa=1}^r
v_{\kappa}{\mathbf a}_{\kappa}\otimes {\mathbf
a}_{\kappa}^{\dagger}$. It is worth noting at this point that if
integrating over $d\mu_{(r)}\left(\left\{{\mathbf a}_\kappa\right\}\right)$
does not introduce correlations between $V_{\alpha\alpha}$, the distribution
function acquires, apart from the asymmetry introduced by the
confining potential, a form reminiscent of a quantum-mechanical
transition amplitude: the Vandermonde determinants play the role of
Slater determinants describing antisymmetric many-fermion wave
functions, and the determinantal form of the transition kernel implies
that the fermions are non-interacting albeit subject to an unusual
single-particle Hamiltonian. Correlations between $V_{\alpha\alpha}$ then imply
interactions between the particles.

We now analyse several limiting cases. If $V$ itself is drawn from a
unitary random matrix ensemble ${\mathcal P}(V)\propto
e^{-\mathrm{Tr}V^2/2X^2}$ (and is thus of rank $r=N$), the integration
over $d\mu_{(N)}\left(\left\{{\mathbf a}_\kappa\right\}\right)$
is subsumed into the integration over $V$. In ``Cartesian''
coordinates
$\prod_\alpha dV_{\alpha\alpha}\prod_{\alpha<\beta}d\Re V_{\alpha\beta}d\Im V_{\alpha\beta}$ the
matrix elements $V_{\alpha\alpha}$ {\em are} explicitly uncorrelated and
independently Gaussian distributed. The transition kernel is thus
proportional to $\det_{\alpha\beta}e^{-X^2
\left(\epsilon_\alpha-\epsilon'_\beta\right)^2/2}$, corresponding to free
propagation of the ordinary one-dimensional fermions in the
quantum-mechanical analogy (with $X^2$ playing the role of imaginary
time). An identical joint distribution arises also in the study of
one-dimensional non-intersecting random walkers~\cite{nagao}.

Returning to the case of {\em fixed} $V$, and still considering $r\sim
N$, we note that the $N^2-N$ ``gauge degrees of freedom'' of $V$ enter
only through $N$ combinations $V_{\alpha\alpha}$. An application of the central
limit theorem then implies that, after integrating the ``gauge degrees
of freedom'' out, $V_{\alpha\alpha}$ become independently distributed with the
Gaussian weight $\exp\left\{-N^2 V_{\alpha\alpha}^2/2 \mathrm{Tr}
V^2\right\}$. Consequently, the form of the kernel coincides with the
case when $V$ is random. 


Let us turn now to the case when $r\sim O(1)$. 
The remaining step in the derivation of the joint distribution
function of the eigenvalues is to implement the integration over
$d\mu_{(r)}\left(\left\{\mathbf{a}_\kappa\right\}\right)$. This can be
done in two ways which lead to different forms of the distribution
functions but are ultimately equivalent in the $N\to\infty$ limit when
applied to the calculation of the correlation functions. The first --
exact -- procedure requires enforcing the conditions of orthonormality
between the $r$ complex vectors $\mathbf{a}_\kappa$. Although quite
simple in the $r=1$ case, already for $r=2$ it leads to a somewhat
lengthy calculation, as we will show below, and it quickly becomes
unmanageable for larger values of $r$. If, however, the ultimate
interest is in the correlation functions $R_{nm}$ in which the scaling
limit $N\to\infty$ is taken at {\em fixed} values of $n$ and $m$,
integrating out the remaining $2N-n-m$ ``energies'' leaves the
correlation function dependent only on a finite number of components
$a_{\kappa\alpha}$ of the $r$ vectors $\mathbf{a}_\kappa$. According to a
well-known result in RMT \cite{mehta}, the latter are distributed
independently according to the Porter-Thomas formula
$e^{-N|a_{\kappa\alpha}|^2}$. The distribution functions obtained by
replacing the {\em whole} of
$d\mu_{(r)}\left(\left\{\mathbf{a}_\kappa\right\}\right)$ with the
product of independent Porter-Thomas distributions for each of the
$Nr$ components of $\mathbf{a}_\kappa$ can be interpreted as resulting
from the substitution of a ``canonical'' ensemble for the gas of
levels in place of the ``micro-canonical'' ensemble implied by the
identity $\mathrm{Tr}H'=\mathrm{Tr}\left(H+V\right)$.

\subsection{Exact (``micro-canonical'') joint distribution functions} 

To illustrate the method, let us first consider the $r=1$ case. The
corresponding distribution function and the lowest-order nontrivial
correlation function ($R_{11}$ in our notation) have been derived
using a somewhat different approach in Ref. \cite{aleiner98}. $V$ is
parametrised as $vN\mathbf{a}\otimes\mathbf{a}^\dagger$, and for a
single normalised complex vector $\mathbf{a}$ the measure
$d\mu_{(1)}\left(\mathbf{a}\right)$ becomes
\begin{equation}
d\mu_{(1)}\left(\mathbf{a}\right)=J_1\delta
\left(\mathbf{a}^\dagger\mathbf{a}-1\right)
e^{-\eta\mathbf{a}^\dagger\mathbf{a}}d\left[\mathbf{a}\right]
\end{equation}
where $d\left[\mathbf{a}\right]$ denotes the unconstrained measure
$\prod_{\alpha=1}^N d a_\alpha d a^*_\alpha/\pi$, and $\eta$
is a positive infinitesimal introduced for convergence. The
normalisation constant $J_1$ is determined from
\begin{equation}
\fl 1=J_1\int d\left[\mathbf{a}\right] \int_{-\infty}^\infty
\frac{dx}{2\pi} e^{ix-i(x-i\eta)\mathbf{a}^\dagger\mathbf{a}} 
=J_1 \int_{-\infty}^\infty\frac{dx}{2\pi}\frac{e^{ix}}
{\left(ix+\eta\right)^N}=\frac{J_1}{\left(N-1\right)!}.
\end{equation}
Substituting this measure into (\ref{genericp}) we obtain
\begin{eqnarray}
\mathcal{P}\left(\{\epsilon\},\{\epsilon'\}\right)=
\frac{C_\mathrm{U}\left|c\right|^2\mu_N}
{\left(2\pi\right)^{N^2-N}}e^{-\sum_{\alpha=1}^NW\left(\epsilon_\alpha\right)}\Delta\left(H\right)
\Delta\left(H'\right)\frac{(N-1)!}{(N|v|)^{N-1}}  \nonumber \\
\times\delta\left(\sum_\beta\epsilon'_\beta-\sum_\alpha\epsilon_\alpha-Nv\right)
\det_{\alpha\beta}\theta\left[\mathrm{Sign}\, v\left(\epsilon'_\beta-\epsilon_\alpha\right)\right].
\end{eqnarray}
As expected \cite{aleiner98}, the determinant is nonzero only if the
two sets of levels satisfy the ``interleaved combs'' constraint
$\epsilon_\alpha<\epsilon'_\alpha<\epsilon_{\alpha+1}$ for $v>0$ and
$\epsilon_{\alpha-1}<\epsilon'_\alpha<\epsilon_\alpha$ for $v<0$. 


We next consider the $r=2$ case.  It is convenient to use a
symmetrised expression $V= vN\left({\mathbf a}\otimes{\mathbf
a}^{\dagger}-{\mathbf b}\otimes{\mathbf b}^{\dagger}\right)$ where
${\mathbf a}$ and ${\mathbf b}$ are two complex mutually orthogonal
$N$-component vectors of length $1$, and $v$ is the scalar (positive)
parameter characterising the magnitude of $V$. More general
expressions can be reduced to this form by a simple rescaling in the
final results. It is also convenient to re-exponentiate the
$\delta$-functions in the determinant in (\ref{genericp}) so that the
$\lambda$ integrals are performed at a later stage in the calculation.
The measure $d\mu_{(2)}\left(\mathbf{a},\mathbf{b}\right)$ is
explicitly parametrised as
\begin{equation}
d\mu_{(2)}\left(\mathbf{a},\mathbf{b}\right)=J_2d\left[{\mathbf
a}\right]d\left[{\mathbf b}\right]\delta\left({\mathbf a},{\mathbf b}\right)
e^{-\eta\left(\mathbf{a}^\dagger\mathbf{a}+\mathbf{b}^\dagger\mathbf{b}\right)}
\end{equation}
where $\delta\left({\mathbf a},{\mathbf b}\right)$ stands for 
\begin{equation}
\delta\left({\mathbf
a}^{\dagger}{\mathbf a}-1\right)\delta\left({\mathbf
b}^{\dagger}{\mathbf b}-1\right) \delta\left({\mathbf
a}^{\dagger}{\mathbf b}+{\mathbf
b}^{\dagger}{\mathbf a}\right)\delta\left(\frac{1}{i}\left[{\mathbf
a}^{\dagger}{\mathbf b}-{\mathbf
b}^{\dagger}{\mathbf a}\right]\right),
\end{equation}
and $J_2$ is the normalisation constant. To compute $J_2$, the
delta-functions are exponentiated with the help of two real Lagrange
multipliers $x$ and $y$, and a complex Lagrange multiplier $z$: 
\begin{eqnarray*}
\fl 1=J_2\int\left[d{\mathbf a}\right]\left[d{\mathbf b}\right]
\delta\left({\mathbf a},{\mathbf b}\right)
\exp\left\{-\eta\left(\mathbf{a}^\dagger\mathbf{a}+\mathbf{b}^\dagger\mathbf{b}\right)\right\}
 =\! J_2 \! \int\frac{dx dy d\Re z
d\Im z}{\left(2\pi\right)^4} \left[d{\mathbf
a}\right]\left[d{\mathbf b}\right]  \\
\times\exp\left\{i\left(x-i\eta\right)\left(1-{\mathbf a}^{\dagger}{\mathbf
a}\right)\! +i\left(y-i\eta\right)\left(1-{\mathbf b}^{\dagger}{\mathbf
b}\right)\! -iz{\mathbf a}^{\dagger}{\mathbf b}\! -iz^*{\mathbf
b}^{\dagger}{\mathbf a}\right\}.
\end{eqnarray*}
Performing the integrals, we find
\begin{equation}
\label{j}
J_2=4\pi\left(\frac{2}{\pi}\right)^{2N}\left(-1\right)^N\left(N-1\right)!
\left(N-2\right)!
\end{equation}
With these definitions, we have
\begin{eqnarray}
\label{9}
\fl {\mathcal P}\left(\left\{\epsilon\right\},
\left\{\epsilon^{\prime}\right\}\right) 
=\frac{C_{\mathrm{U}}J_2}{\left(2\pi\right)^{N^2}}
\Delta\left(H\right)
\Delta\left(H'\right) \int d\mu\left(U_{\lambda}\right)\prod_{k=1}^N
d\lambda_k
d\left[{\mathbf a}\right]d\left[{\mathbf b}\right]
\delta\left({\mathbf a},{\mathbf b}\right)
\exp\left\{i\mathrm{Tr}\left(\lambda V\right)\right\} \nonumber \\
\fl \times 
\sum_{\sigma,\sigma^{\prime}}\left(-1\right)^{\sigma+\sigma'}
\exp\left\{i\sum_{\alpha=1}^N\left(\epsilon_{\alpha}\lambda_{\sigma\left(\alpha\right)}-\epsilon^{\prime}_{\alpha}
\lambda_{\sigma'\left(\alpha\right)}\right)\right\}
\exp \left\{-\sum_{\alpha=1}^NW\left(\epsilon_\alpha\right)\right\}.
\end{eqnarray}
Proceeding as in the computation of $J_2$, we obtain
\begin{eqnarray}
\label{14}
\fl {\mathcal P}\left(\left\{\epsilon\right\},
\left\{\epsilon^{\prime}\right\}\right) = 
\frac{C_{\mathrm U}\left|J_2\right|\left|c\right|^2\mu_N}{\left(2\pi\right)^{N^2}}
\Delta\left(H\right) \Delta\left(H'\right)
\left(\frac{\pi^2}{Nv}\right)^N\int \frac{ dx dy d\Re z
d\Im z}{\left(2\pi\right)^4} e^{i\left(x+y\right)} \nonumber \\
\fl \times
\sum_{\sigma,\sigma^{\prime}} \left(-1\right)^{\sigma+\sigma'}
\prod_{k=1}^N \int d\lambda_k\frac{\exp\left\{i\lambda_k
\left(\epsilon_{\sigma\left(k\right)}-\epsilon^{\prime}_{\sigma^{\prime}
\left(k\right)}\right)/Nv
\right\}}
{\left(\lambda_k-x+i\eta\right)\left(\lambda_k+y-i\eta\right)
+\left|z\right|^2}
\exp\left\{-\sum_{\alpha=1}^NW\left(\epsilon_\alpha\right)\right\},
\end{eqnarray}
After integrating out the eigenvalues of $\lambda$ we obtain
\begin{eqnarray}
\label{int1}
\fl {\mathcal P}\left(\left\{\epsilon\right\},
\left\{\epsilon^{\prime}\right\}\right) = 
\frac{C_{\mathrm{U}}\left|J_2\right|\left|c\right|^2\mu_N N!}
{\left(2\pi\right)^{N^2-N}}
\Delta\left(H\right) \Delta\left(H'\right)
\left(\frac{\pi^2}{Nv}\right)^N\frac{1}{8\left(2\pi\right)^3}
e^{-\sum_{\alpha=1}^NW\left(\epsilon_\alpha\right)} \nonumber \\
\fl \times \sum_{\sigma}\left(-1\right)^\sigma
\int_{-\infty}^{\infty}
d\tau \exp\left\{\frac{i\tau\sum_{\alpha=1}^N
\left(\epsilon_{\alpha}-\epsilon^{\prime}_{\sigma
\left(\alpha\right)}\right)}{2Nv}\right\}
\int_{-\infty}^{\infty}dte^{it}\int_{0}^{\infty}\zeta d\zeta
\nonumber \\
\fl \left\{\theta\left(t^2-\zeta^2\right)\left(-is_t\right)^N
\frac{\exp\left\{-i\sqrt{t^2-\zeta^2}\,
s_t \Delta_{\sigma}\right\}}
{\left(\sqrt{t^2-\zeta^2}-i\eta s_t\right)^N}
+
\theta\left(\zeta^2-t^2\right)\frac{\exp\left\{-\sqrt{\zeta^2-t^2}\,
\Delta_{\sigma}\right\}}{\left(\sqrt{\zeta^2-t^2}+
i\eta s_t\right)^N}\right\}, \nonumber
\end{eqnarray}
where $\tau=x-y$, $t=x+y$, $\zeta=2\left|z\right|$,
$s_t\equiv\mathrm{Sign}\,t$, the ambiguities in the analytic
properties of the integrand are resolved by letting $t\rightarrow
t-i\eta$, and $\Delta_{\sigma}=\sum_{\alpha=1}^N
\left|\epsilon_{\alpha}-\epsilon^{\prime}_{\sigma
\left(\alpha\right)}\right|/2Nv$. The extra factor $N!$ comes from
performing one of the sums over permutations. Note that the parameter
$\Delta_{\sigma}$ depends only on the absolute values of the distances
between the levels in contrast to the case of rank $r=1$ perturbation
\cite{aleiner98}.

Integration over $\tau$ gives a factor of $4\pi
Nv\,\delta\left[\sum_\alpha\left(
\epsilon_{\alpha} -\epsilon^{\prime}_{\alpha}\right)\right]$ 
which simply enforces the identity $\mathrm{Tr}\,H=\mathrm{Tr}\, H'$. 
Performing the remaining integrations, we find
\begin{eqnarray}
\label{result}
\fl {\mathcal P}\left(\left\{\epsilon\right\},
\left\{\epsilon^{\prime}\right\}\right) =
\frac{2C_{\mathrm{U}}\left|c\right|^2\mu_N}{\left(2\pi\right)^{N^2-N}}
\left(\frac{4}{Nv}\right)^{N-1}N!\left(N-1\right)!\,
\delta\left[\sum_\alpha\left(
\epsilon_{\alpha} -\epsilon^{\prime}_{\alpha}\right)\right] \nonumber \\
\times
\Delta\left(H\right) \Delta\left(H'\right)
e^{-\sum_{\alpha=1}^NW\left(\epsilon_\alpha\right)}
\sum_{\sigma}\left(-1\right)^\sigma
\theta\left(1-\Delta_{\sigma}\right)
\left(1-\Delta_{\sigma}\right)^{N-2}.
\end{eqnarray}
Note the hard constraint on the sum of the absolute values of the
level shifts
$\sum_\alpha\left|\epsilon_\alpha-\epsilon'_{\sigma\left(\alpha\right)}\right|\leq
2Nv$. Note also that this result cannot be simplified in an obvious
way in the $N\to\infty$ limit since $\Delta_\sigma$ {\em does not}
scale as $1/N$. Indeed, the natural scale for $v$ is the mean level
spacing $\bar{\Delta}$, and
$\sum_\alpha\left|\epsilon_\alpha-\epsilon'_{\sigma\left(\alpha\right)}\right|\sim
O(N\bar{\Delta})$ so that the $\Delta_\sigma\sim O(1)$. The
distribution functions (\ref{genericp}) and (\ref{detdist})
below are equivalent only in the sense that they lead to identical
correlation functions in the ``thermodynamic'' limit $N\to\infty$.



\subsection{``Canonical'' joint distributions} 

As discussed above, after neglecting $O(1/N)$ corrections, the measure
$d\mu_{(r)}$ for finite $r$ is brought to the simple form
\begin{equation}
d\mu_{(r)}\left(\left\{\mathbf{a}_\kappa\right\}\right)=
\prod_{\kappa=1}^r\prod_{\alpha=1}^N\frac{Nda_{\kappa
    \alpha}da^*_{\kappa \alpha}}{\pi}
e^{-N\left|a_{\kappa\alpha}\right|^2}.
\end{equation}
Due to the factorised structure of this measure, the transition
kernel in (\ref{genericp}) acquires the determinantal form
\begin{equation}
\label{detdist}
\det_{\alpha\beta}
\hat{\mathcal{D}}^{-1}_{\left\{\epsilon_\alpha;\hat{v}\right\}}
\delta(\epsilon_\alpha-\epsilon_\beta)
\end{equation}
where the action of the differential operator $\hat{\mathcal{D}}_{\left\{\epsilon;\hat{v}\right\}}$ on an
arbitrary function $f(\epsilon)$ is defined as
\begin{equation}
\hat{\mathcal{D}}_{\left\{\epsilon;\hat{v}\right\}}
f(\epsilon)=\det\left(\hat{1}-\hat{v}d/d\epsilon\right)f\left(\epsilon\right)
\end{equation}
and its inverse has a convenient integral representation
\begin{equation}
\hat{\mathcal{D}}^{-1}_{\left\{\epsilon;\hat{v}\right\}}
f(\epsilon)=\int\prod_{\kappa=1}^r\frac{d\psi_\kappa
d\psi^*_\kappa}{\pi}\exp\left\{-\left|\boldsymbol{\psi}\right|^2\right\}
f(\epsilon+\boldsymbol{\psi}^*\hat{v}\boldsymbol{\psi}).
\end{equation}

In the opposite case of infinite rank, the central limit theorem
ensures that the diagonal matrix elements $V_{\alpha\alpha}$ are independently
Gaussian distributed. In consequence, the differential operator
$\hat{\mathcal{D}}_{\left\{\epsilon;\hat{v}\right\}}$
simplifies to
\begin{equation}
\label{dinfrank}
\hat{\mathcal{D}}_{\left\{\epsilon;\hat{v}\right\}}
=\exp\left\{-\frac{1}{2}\mathrm{Tr}\, \hat{v}^2
\frac{d^2}{d\epsilon^2}\right\}.
\end{equation}
Here and below when considering the infinite rank case we assume
$\mathrm{Tr}\,V=0$ since a non-zero value of $\mathrm{Tr}\,V$ can be
accounted for by a trivial uniform shift of the spectrum.

\section{Parametric correlation functions}

\subsection{Parametric correlation functions in the energy representation}

The objective of this section is to present a derivation of the
multipoint correlation function $R_{nm}$ [equations (\ref{corrdet})
and (\ref{corrdet2}) below]. The joint distribution function defined
by equations (\ref{genericp}) and (\ref{detdist}) serves as the basis
for this calculation which employs a scheme based on the orthogonal
polynomial expansion. In accordance with the definition (\ref{rnm}),
the parametric correlation function is represented as
\begin{equation}
\fl
R_{nm}\left(\left\{E_a\right\}_{a=1}^n,\left\{E'_b\right\}_{b=1}^m\right)=
\int\prod_{\alpha=1}^Nd\epsilon_\alpha\prod_{\beta=1}^Nd\epsilon'_\beta
\mathcal{P}\left(\left\{\epsilon\right\},\left\{\epsilon'\right\}\right)S(\{E\})S'(\{E'\}),
\end{equation}
where the source terms $S(\{E\})$ and $S'(\{E'\})$ are defined as
\begin{equation}
\fl \qquad S(\{E\})=\prod_{a=1}^n\sum_{\alpha_a=1}^N
\delta\left(E_a-\epsilon_{\alpha_a}\right),\quad
S'(\{E'\})=\prod_{b=1}^m\sum_{\beta_b=1}^N\delta\left(E'_b-\epsilon'_{\beta_b}\right).
\end{equation}
Integrating over all $\epsilon'$ we obtain
\begin{eqnarray}
R_{nm}\left(\left\{E_a\right\}_{a=1}^n,\left\{E'_b\right\}_{b=1}^m\right)=
\int\prod_{\alpha=1}^Nd\epsilon_\alpha \Delta(\epsilon)
\exp\left\{-\sum_{\alpha=1}^N W\left(\epsilon_\alpha\right)\right\}S(\{E\}) \nonumber \\
\ns \times\prod_{\beta=1}^N\hat{\mathcal{D}}^{-1}_{\left\{\epsilon_\beta;\hat{v}\right\}}\left[\Delta(\epsilon)
  \prod_{b=1}^m\sum_{\beta_b=1}^N\delta(E'_b-\epsilon_{\beta_b})\right].
\end{eqnarray}
Let us denote as $\pi_n$, and  $\pi_m$, respectively, 
the set of all {\em distinct} indices $\alpha_a$ in
the multiple sum over $\alpha_1,\dots,\alpha_n$, and
the set of all distinct indices $\beta_b$ in the multiple sum over
$\beta_1,\dots,\beta_m$. It is now convenient to split the product of
$\hat{\mathcal{D}}^{-1}$'s as 
\begin{equation}
\prod_{\beta\in\pi_m}\hat{\mathcal{D}}^{-1}_{\left\{\epsilon_\beta;\hat{v}\right\}}
\prod_{\beta=1,\beta\notin\pi_m}^N\hat{\mathcal{D}}^{-1}_{\left\{\epsilon_\beta;\hat{v}\right\}}. \nonumber
\end{equation}
All the factors in the second product act only on the variables in
$\Delta(\epsilon)$ while the operators in the first product act on the
variables in both $\Delta$ and the source term delta-functions.
The Vandermonde determinant involves only polynomials of degree $p-1$
in its $p$-th row, so that the action of a product of $N$ identical
differential operators transforms the elements in its $p$-th row into
different polynomials of the same degree and with the same leading
term. Consequently,
\begin{equation}
\fl \prod_{\beta=1}^N\hat{\mathcal{D}}^{-1}_{\left\{\epsilon_\beta;\hat{v}\right\}}\Delta(\epsilon)\equiv\Delta(\epsilon),
\quad\mathrm{and}\quad
\prod_{\beta=1,\beta\notin\pi_m}^N\hat{\mathcal{D}}^{-1}_{\left\{\epsilon_\beta;\hat{v}\right\}}\Delta(\epsilon)=
\prod_{\beta\in\pi_m}\hat{\mathcal{D}}_{\left\{\epsilon_\beta;\hat{v}\right\}}\Delta(\epsilon).
\end{equation}
Integrating by parts over the variables in $\pi_m$ we arrive at
\begin{eqnarray}
\label{intermed}
\fl R_{nm}\left(\left\{E_a\right\}_{a=1}^n,\left\{E'_b\right\}_{b=1}^m\right)=
\sum_{\substack{\beta_1,\dots,\beta_m \cr \alpha_1,\dots,\alpha_n}}
\int\prod_{\alpha=1}^Nd\epsilon_\alpha
\left[\prod_{\beta\in\pi_m}\hat{\mathcal{D}}_{\left\{\epsilon_\beta;\hat{v}\right\}}\Delta(\epsilon)\right]
\prod_{b=1}^m\delta\left(E'_b-\epsilon_{\beta_b}\right)
\nonumber \\
\times
\prod_{\beta\in\pi_m}\hat{\mathcal{D}}^{\dagger {}^{-1}}_{\left\{\epsilon_\beta;\hat{v}\right\}}
\left[\exp\left\{-\sum_{\alpha=1}^N W\left(\epsilon_\alpha\right)\right\}
\Delta(\epsilon)\prod_{a=1}^n\delta\left(E_a-\epsilon_{\alpha_a}\right)\right],
\end{eqnarray}
where
$\hat{\mathcal{D}}^{\dagger}_{\left\{\epsilon;\hat{v}\right\}}=\hat{\mathcal{D}}_{\left\{\epsilon;-\hat{v}\right\}}$. 
The above expression is manifestly asymmetric with respect to the
confining potential $W(\epsilon)$. As long as all the energies are
close to the centre of the band, and the change in the average density
of states induced by the potential $V$ is negligible, the action of
$\hat{\mathcal{D}}$ on $W$ can be ignored. It is instructive, however,
to consider a more general case. To this end, we introduce a new
operator
\begin{equation}
\hat{\widetilde{\mathcal{D}}}_{\left\{\epsilon;\hat{v}\right\}}=\exp\left\{-W\left(\epsilon\right)/2\right\}
\hat{\mathcal{D}}_{\left\{\epsilon;\hat{v}\right\}}\exp\left\{W\left(\epsilon\right)/2\right\}.
\end{equation}
When (\ref{intermed}) is rewritten in terms of
$\hat{\widetilde{\mathcal{D}}}$ and
$\hat{\widetilde{\mathcal{D}}}^\dagger$, each of the Vandermonde
determinants appears in combination with
$\prod_{\alpha=1}^N\exp\left\{-W\left(\epsilon_\alpha\right)/2\right\}$,
so that they can be equivalently written as
$\det_{\alpha\beta}\varphi_{\alpha -1}(\epsilon_\beta)$ where
$\varphi_{\alpha
-1}(\epsilon)=\exp\left\{-W\left(\epsilon\right)/2\right\} p_{\alpha
-1}(\epsilon)$, and $p_\alpha\left(\epsilon\right)$ are the
polynomials orthogonal with respect to the weight
$\exp\left\{-W(\epsilon)\right\}$.  Expanding the determinants, and
using the orthonormality of the set of functions
$\varphi_{\alpha}(\epsilon)$, we arrive at the following
representation of the correlation function as an $(n+m)\times (n+m)$
determinant:
\begin{eqnarray}
\label{corrdet}
\fl \quad\quad R_{nm}\left(\left\{E_a\right\}_{a=1}^n,\left\{E'_b\right\}_{b=1}^m\right)
\nonumber \\
\fl \quad\quad=\det\left(\begin{array}{cc} \mathrm{k}(E_{a_1},E_{a_2}) & 
  \hat{\tilde{\mathcal{D}}}^{-1}_{\left\{E'_{b_2};\hat{v}\right\}}
\left[\mathrm{k}(E_{a_1},E'_{b_2})-\delta\left(E_{a_1}-E'_{b_2}\right)\right]
 \\ \hat{\tilde{\mathcal{D}}}^{\dagger}_{\left\{E'_{b_1};\hat{v}\right\}}
\mathrm{k}(E'_{b_1},E_{a_2}) & \hat{\tilde{\mathcal{D}}}^{-1}_{\left\{E'_{b_2};\hat{v}\right\}}
\hat{\tilde{\mathcal{D}}}^{\dagger}_{\left\{E'_{b_1};\hat{v}\right\}} 
\mathrm{k}(E'_{b_1},E'_{b_2})
\end{array}\right),
\end{eqnarray}
where the indices $a$ and $a'$ ($b$ and $b'$) run from $1$ to $n$
(from $1$ to $m$). The kernel
\begin{equation}
\label{kee}
\mathrm{k}(E,E')=\sum_{p=0}^{N-1}\varphi_p(E)\varphi_p(E')
\end{equation}
describes the correlations among the levels of the original matrices
$H$, while the lower right corner of the matrix describes the
correlation functions {\em within} the ensemble defined by the
distribution function
$\mathcal{P}\left(H\right)=C_{\mathrm{U}}\exp\left\{-\mathrm{Tr}\,W\left(H-V\right)\right\}$. The
cross-correlations between $H$ and $H'=H+V$ are encoded in the
off-diagonal blocks of (\ref{corrdet}). In the bulk scaling limit the
kernel $\mathrm{k}$ takes the form \cite{BZ} 
\begin{equation}
\label{scaledk}
\mathrm{k}\left(E,E'\right)=\Im
\exp\left\{i \pi \omega
\rho\left(\epsilon\right)\right\}/\omega\equiv\pi\rho\left(\epsilon\right)
k\left[\pi\rho\left(\epsilon\right)\omega\right], 
\end{equation}
where $\omega=E'-E$ and $\epsilon=\left(E'+E\right)/2$.

The bulk of the present paper is devoted to the case when the strength
of the perturbation $V$ is such that the parametric correlators
between the ``old'' and the ``new'' levels are $O(1)$. This condition
is fulfilled if $\mathrm{Tr} V^2\sim O(N^2\Delta^2)$. In this regime
the operator $\hat{\tilde{\mathcal{D}}}_{\left\{\epsilon;\hat{v}\right\}}$
can be explicitly evaluated using a representation of
$\hat{\mathcal{D}}$ as an integral over a vector of Grassmann
variables $\boldsymbol{\chi}$:
\begin{eqnarray}
\fl \hat{\tilde{\mathcal{D}}}_{\left\{\epsilon;\hat{v}\right\}}\equiv
\exp\left[-W\left(\epsilon\right)/2\right]
\int d\boldsymbol{\chi}
d\boldsymbol{\chi}^* \exp\left\{-\left|\boldsymbol{\chi}\right|^2
-\boldsymbol{\chi}^*\hat{v}\boldsymbol{\chi}\frac{d}{d\epsilon}\right\} 
\exp\left[W\left(\epsilon\right)/2\right] \nonumber \\
 \approx \int d\boldsymbol{\chi} d\boldsymbol{\chi}^* 
\exp\left[W'\left(\epsilon\right)\boldsymbol{\chi}^*\hat{v}\boldsymbol{\chi}/2\right]
\exp\left\{-\left|\boldsymbol{\chi}\right|^2
-\boldsymbol{\chi}^*\hat{v}\boldsymbol{\chi}\frac{d}{d\epsilon}\right\}
\nonumber \\
 =\det\left[1-\hat{v}W'\left(\epsilon\right)/2\right]
\hat{\mathcal{D}}_{\left\{\epsilon;\hat{\mathfrak{r}}\left(\hat{v}\right)\right\}},
\end{eqnarray}
where
$\hat{\mathfrak{r}}\left(\hat{v}\right)=\hat{v}/\left(1-\hat{v}W'\left(\epsilon\right)/2\right)$.
Assuming that all the energies in both sets $\{E\}$ and $\{E'\}$ lie
within a few level spacings of some central energy $\epsilon$, and
since the determinantal structure of (\ref{corrdet}) ensures that the
operators $\hat{\mathcal{D}}$ and $\hat{\mathcal{D}}^{-1}$ come in
pairs, the determinant in the prefactor can be dropped.

The derivative of the confining potential is related to the real part
of the {\em average} Green function via
\begin{equation}
\label{wprime}
W'\left(\epsilon\right)=2\dashint
\frac{\rho\left(z\right)}{z-\epsilon}\equiv 2\Re
G\left(\epsilon\right).
\end{equation}
Consequently, $\mathfrak{r}\left(v\right)$ can be represented as the
solution of the equation
\begin{equation}
\label{reactance}
\hat{\mathfrak{r}}\left(\hat{v}\right)=\hat{v}+\hat{v}\Re
G\left(\epsilon\right) \hat{\mathfrak{r}}\left(\hat{v}\right),
\end{equation}
which can be immediately identified as the equation for the {\em
reactance matrix} for scattering off the potential $V$. In fact, as
can be shown using the methods of statistical field theory \cite{MSS},
the parametrisation of $\hat{\mathcal{D}}$ in terms of the
corresponding reactance matrix retains its validity in the more
general case of {\em non-invariant} distributions
$\mathcal{P}\left(H\right)$. The eigenvalues of the reactance matrix
are conveniently parametrised as $\tan\delta_\kappa$ where
$\delta_\kappa$ are the corresponding phase shifts.

As discussed above, the parametric correlations generically fall into
two regimes corresponding to the rank of $V$ being finite or
infinite. In the latter case, the condition $\mathrm{Tr} V^2\sim
O(N^2\Delta^2)$ ensures that, for a generic eigenvalue $v_i$ of $\hat{v}$,
and at a generic position $\epsilon$ in the spectrum,
$v_iW'(\epsilon)\ll 1$. Thus the distinction between
$\hat{\tilde{\mathcal{D}}}$ and $\hat{\mathcal{D}}$ is meaningful only
in the finite rank case, and, to simplify notations, we will
use the symbol $\hat{\mathcal{D}}$ in both cases with its
parametrisation implied by the context.

Similarly, to the leading order in the $1/N$ expansion, the density of
states $\rho\left(\epsilon\right)$ can be approximated by an
$\epsilon$-independent inverse mean level spacing $1/\bar{\Delta}$,
thus simplifying (\ref{corrdet}) to the form quoted in
\cite{SMS}:
\begin{eqnarray}
\label{corrdet2}
\fl \quad \mathfrak{R}_{nm}\left(\left\{E_a\right\}_{a=1}^n,\left\{E'_b\right\}_{b=1}^m\right)
\nonumber \\
\fl \quad =\det\left(\begin{array}{cc} k\left(\varepsilon_{a_2}-\varepsilon_{a_1}\right) & 
  \hat{\mathcal{D}}^{-1}_{\left\{\varepsilon'_{b_2};\hat{\mathrm{v}}\right\}}
\left(k\left[\varepsilon'_{b_2}-\varepsilon_{a_1}\right]-
\delta\left[\varepsilon'_{b_2}-\varepsilon_{a_1}\right]\right)  \\ 
\hat{\mathcal{D}}_{\left\{\varepsilon_{a_2};\hat{\mathrm{v}}\right\}}
k\left(\varepsilon_{a_2}-\varepsilon'_{b_1}\right) &  
k\left(\varepsilon'_{b_2}-\varepsilon'_{b_1}\right)
\end{array}\right),
\end{eqnarray}
where $\hat{\mathrm{v}}=\hat{v}/\bar{\Delta}$, and
$\mathfrak{R}_{nm}=\bar{\Delta}^{n+m} R_{nm}$. In the $r\to\infty$
limit this equation reproduces the results obtained previously
\cite{SA,frelkanzieper}. The representation of $R_{nm}$ in the form of
$\lambda$ integrals characteristic of the non-linear sigma model
calculations \cite{SA,MSS} corresponds to the Fourier decomposition of
$\hat{\mathcal{D}}k$ and $\hat{\mathcal{D}}^{-1}[k-\delta]$ in
(\ref{corrdet2}).

\subsection{Universality in shifted distributions}

To draw a connection with some previous studies of parametrically
deformed random matrix ensembles \cite{hikami,kanzieper}, let us
briefly consider the opposite case of a perturbation strong enough to
significantly affect the density of states. In this regime, the
cross-correlations are small as $1/N$. There remains, however, the
question of whether the correlation functions of a matrix ensemble
defined by the shifted distribution
$\mathcal{P}_V\left(H\right)=C_U\exp\left\{-\mathrm{Tr}\,W\left(H-V\right)\right\}$
are universal. To the best of our knowledge, this question has been so
far answered in the affirmative only in the case when $W$ is quadratic
\cite{hikami} (see also \cite{PZinn} where the case of non-Gaussian
potential with a linear source term was considered). The formalism
developed above affords an opportunity to extend the proof to the case
of arbitrary $W(H)$ (still, however, assuming strong confinement). The
corresponding correlation functions, as follows from (\ref{corrdet}),
have the determinantal structure with the kernel
$\mathrm{k}_v(\omega,\epsilon)=\hat{\tilde{\mathcal{D}}}^{-1\dagger}_{\left\{\epsilon_1;\hat{v}\right\}}
\hat{\tilde{\mathcal{D}}}_{\left\{\epsilon_2;\hat{v}\right\}}\mathrm{k}\left(\omega,\epsilon\right)$.
The operator $\hat{\mathcal{D}}$ in this regime is given by equation
(\ref{dinfrank}).
Expanding the difference
$W\left(\epsilon_2\right)-W\left(\epsilon_1\right)\approx \omega
W'\left(\epsilon\right)$, we obtain
\begin{equation}
\label{kv}
\fl \mathrm{k}_v(\omega,\epsilon)=\exp\left\{-\omega W'\left(\epsilon\right)/2\right\}
\exp\left\{-\pi^2 x^2\partial_\omega\partial_\epsilon\right\}
\exp\left\{\omega W'\left(\epsilon\right)/2\right\} \Im
\frac{\exp\left\{i\omega \pi \rho\left(\epsilon\right)\right\}}{\pi \omega},
\end{equation}
where $x=\left(\mathrm{Tr}\hat{v}^2\right)^{1/2}/\pi$. By virtue of (\ref{wprime}), the
last two exponents can be combined to read $\Im\exp\left\{\omega
\mathcal{G}\left(\epsilon\right)\right\}$, where
$\mathcal{G}\left(z\right)$ is defined as the analytic continuation of
the retarded Green function $G(\epsilon)=\dashint dz
\rho\left(z\right)/\left(z-\epsilon\right)+i\pi\rho\left(\epsilon\right)$
onto the complex plane with cuts along the real axis {\em outside} the
support of the density of states $\rho\left(\epsilon\right)$.
By employing an integral representation of the differential
operator
\begin{equation}
\exp\left\{-\pi^2 x^2\partial_\omega\partial_\epsilon\right\}=
\int\frac{d\zeta d\zeta^*}{\pi^3
  x^2}\exp\left\{-\left|\zeta\right|^2/\pi^2 x^2
-\zeta^*\partial_\omega+\zeta\partial_\epsilon\right\},
\end{equation}
and differentiating with respect to $\epsilon$, equation (\ref{kv}) is
brought to the form
\begin{eqnarray}
\partial_\epsilon \left[\exp\left\{\omega
W'\left(\epsilon\right)/2\right\} \mathrm{k}_v(\omega,\epsilon)\right]
\nonumber \\
= 
\int \frac{d\zeta d\zeta^*}{\pi^3 x^2}
\exp\left\{-\left|\zeta\right|^2/\pi^2 x^2\right\} \Im\exp\left\{\left(\omega-\zeta^*\right) 
\mathcal{G}\left(\epsilon+\zeta\right)\right\} \mathcal{G}'\left(\epsilon+\zeta\right).
\end{eqnarray}
After a change of variables $r=\left|\zeta\right|^2$,
$u=e^{i\arg\zeta}$ we note that, since $\mathcal{G}$ is analytic by
construction, the only non-analyticity of the integral over $u$ is due to
the pole generated by the solution $u_0\left(\epsilon\right)$ of the
equation $\pi^2 x^2 \mathcal{G}\left(\epsilon+u\right)+u=0$ (existence
of multiple solutions of this equation would contradict level number
conservation).
We thus find, upon rescaling $u$
and integrating over $r$,
\begin{eqnarray}
\partial_\epsilon \left[\exp\left\{\omega
W'\left(\epsilon\right)/2\right\} \mathrm{k}_v(\omega,\epsilon)\right]=\Im  \oint
du \frac{\exp\left\{\omega\mathcal{G}\left(\epsilon+u\right)
  \right\}}{\pi^2 x^2\mathcal{G}\left(\epsilon+u\right)+u} \mathcal{G}'\left(\epsilon+u\right)
\nonumber \\
=\Im \frac{\exp\left\{-\omega u_0\left(\epsilon\right)/\pi^2 x^2 \right\}
  \mathcal{G}'\left(\epsilon+u_0\left(\epsilon\right)\right)}
{1+\pi^2 x^2\mathcal{G}'\left(\epsilon+u_0\left(\epsilon\right)\right)},
\end{eqnarray}
where the contribution from integrals along the cuts vanishes upon
taking the imaginary part of this expression. Integrating over
$\epsilon$ and using
$\left[1+\pi^2 x^2\mathcal{G}'\left(\epsilon+u_0\left(\epsilon\right)\right)\right]\left(du_0/d\epsilon\right)
=-\pi^2 x^2 \mathcal{G}'\left(\epsilon+u_0\left(\epsilon\right)\right)$,
we find
\begin{equation}
\mathrm{k}_v(\omega,\epsilon)=\exp\left\{-\omega
W'\left(\epsilon\right)/2+\omega\Re u_0\left(\epsilon\right)\right\}
\frac{\sin\omega\rho_v\left(\epsilon\right)}{\omega},
\end{equation}
where the renormalised density of states is
$\rho_v\left(\epsilon\right)=-\Im
u_0\left(\epsilon\right)/\pi^2 x^2$. The exponential prefactors cancel
out due to the determinantal structure of the correlation functions,
thus establishing their universal form.

\subsection{Parametric correlation functions in the number representation}
The ``complementary'' number-dependent correlation functions are
generically expressed as derivatives of the corresponding generating
functions. The archetypal generating function is the probability
$\mathfrak{P}_{nn'}(J,J')$ to find $\mathfrak{n}\left(J\right)=n$
levels in the interval $J$ of the unperturbed sequence {\em and} $n'$
levels in the interval $J'$ of the perturbed sequence, where
$\mathfrak{n}(J)$ is the number of levels in the interval $J$:
\begin{equation}
\label{pnj}
\fl \mathfrak{P}_{nn'}(J,J';\mathfrak{x})\equiv
\left\langle\delta_{n,\mathfrak{n}\left(J\right)}
\delta_{n',\mathfrak{n}\left(J'\right)}\right\rangle
=\frac{\left(-1\right)^{n+n'}}{n!n'!}\sum_{k=n}^\infty
\sum_{k'=n'}^\infty\frac{\left(-1\right)^{k+k'}r_{kk'}}{\left(k-n\right)!
\left(k'-n'\right)!}.
\end{equation}
Here $r_{kk'}$ represents the correlation function $R_{kk'}$
integrated over the direct product of intervals $J^k\otimes J^{\prime
k'}$ with the corresponding measures $d\mathfrak{m}_J$ and
$d\mathfrak{m}_{J'}$. The intervals $J$ and $J'$ may, generically,
consist of an arbitrary number of disjointed segments. Equation
(\ref{pnj}) is a straightforward generalisation of the non-parametric
analogue $P_n(J)$ (see, e.~g., Appendix 7 of Ref. \cite{mehta}). For
later convenience in carrying out the summations, $r_{kk'}$ can be
represented in the form of a fermionic functional integral
\begin{eqnarray}
\label{r}
\fl \quad \quad r_{kk'}=\det K \int{\mathcal D}\Psi{\mathcal D}\bar{\Psi} \left(\int 
d\mathfrak{m}_J(u)\bar{\xi}(u)\xi(u)\right)^k\left(\int d\mathfrak{m}_{J'}(w)\bar{\eta}(w)\eta
(w)\right)^{k'} \nonumber \\
\times \exp\left\{\int du\int dw \bar{\Psi}\left(u\right)K^{-1} 
\left(u,w\right) \Psi\left(w\right)\right\}
\end{eqnarray}
where $\bar{\Psi} = \left(\bar{\xi},\bar{\eta}\right)$ is a fermionic
doublet. The matrix kernel $\hat{K}$ is easily reconstructed from
(\ref{corrdet}):
\begin{equation}
\label{kmatrix}
K\left(u,w\right)=\left(\begin{array}{cc}\mathrm{k}\left(u,w\right) & \hat{\mathcal
D}^{-1}_w\left[\mathrm{k}\left(u,w\right)-\mathbb{I}\right] \\
\hat{\mathcal D}^{\dagger}_u \mathrm{k}\left(u,w\right) & \hat{\mathcal
D}^{-1}_w\hat{\mathcal D}^{\dagger}_u\mathrm{k}\left(u,w\right)
\end{array} \right),
\end{equation}
the matrix elements of the scalar kernel $\hat{\mathrm{k}}$ are given by
(\ref{kee}), and $\mathbb{I}$ is adopted as a shorthand for the
Dirac $\delta$-function. For brevity we also drop the subscript
$\hat{\mathfrak{r}}\left(\hat{v}\right)$, and $\hat{\mathcal{D}}$ is
understood to be the function of either the single parameter
$\mathrm{Tr}\,\hat{v}^2$ or the $r$ phase shifts $\delta_\kappa$
parameterising $\hat{\mathfrak{r}}\left(\hat{v}\right)$ in the
infinite, and finite rank cases, respectively. 

Equations (\ref{pnj}), (\ref{r}), and (\ref{kmatrix}) can be used directly
to compute a certain class of correlation functions exemplified,
e.~g., by the correlation between level spacings of the original and
perturbed level sequences centred at given {\em energies} $E$, and
$E'$, respectively. Since our main interest, however, is in the
correlation functions in the number representation, we need to be able
to relate the levels $\epsilon'_\beta$ in the new sequence to the
parametric ``descendant'' of a given level $\epsilon_\alpha$ in the
unperturbed sequence. The absence of level crossings allows to do
this by demanding, e.g., that if an interval
$J_\epsilon=\left(-\infty,\epsilon\right]$ of the unperturbed sequence
contains exactly $n$ levels (as counted from the bottom of the band),
the corresponding interval $J'_{\epsilon+\omega}$ of the perturbed
sequence contains exactly $n+q$ levels, where $q$ is fixed and $n$ is
arbitrary. By construction, then, the uppermost level in the interval
$J'_{\epsilon+\omega}$ is separated by $q-1$ levels from the parametric
descendant of the uppermost level in the interval $J_\epsilon$.
Summing over all $n$ we find the corresponding probability
as
\begin{eqnarray}
\label{pjj}
\fl \mathfrak{P}_q(J_\epsilon,J'_{\epsilon+\omega};\mathfrak{x})\equiv
\left\langle\delta_{\mathfrak{n}\left(J_\epsilon\right),\mathfrak{n}\left(J'_{\epsilon+\omega}\right)+q}\right\rangle
=\sum_{n=0}^\infty\frac{1}{n!(n+q)!}  \!\!\sum_{\begin{subarray}{c}
k=n \cr k'=n+q\end{subarray}}^\infty
\frac{\left(-1\right)^{k+k'+q}r_{kk'}}{\left(k-n\right)!
\left(k'-n-q\right)!}  \nonumber \\
=\int_0^{2\pi}\!\!\!\frac{d\phi}{2\pi} e^{iq\phi}
\sum_{\begin{subarray}{c}k=0 \cr k'=q\end{subarray}}^{\infty}
\frac{\left(-1\right)^{k+k'+q}r_{kk'}}{k!k'!}  z_{+}^kz_{-}^{k'},
\end{eqnarray}
where $z_{\pm}=1+e^{\pm i\phi}$. Substituting (\ref{r}) into
(\ref{pjj}), one finds
\begin{equation}
\label{det1}
\mathfrak{P}_q\left(J,J';\mathfrak{x}\right)=\left(-1\right)^q
\int_0^{2\pi}\frac{d\phi}{2\pi}e^{iq\phi}
Z_q\left(\omega,\mathfrak{x};\phi\right),
\end{equation}
where
\begin{equation}
\fl Z_q\left(\omega,\mathfrak{x};\phi\right)=
\det\left[\mathbb{I} \sigma_0-
\hat{K}\Pi\left(\phi\right)\right] 
-\left.
\sum_{k'=0}^{q-1}\frac{\partial_\gamma^{k'}}{k'!}
\det\left[\mathbb{I}\sigma_0-\hat{K}\Pi_{\gamma}\left(\phi\right)\right]
\right|_{\gamma=0},
\end{equation}
and, denoting the Pauli matrices as $\sigma_i$,
\begin{eqnarray*}
\Pi\left(\phi\right)=\left(\begin{array}{cc} z_{+} & 0 \\ 0 &  z_{-}
\end{array}\right),\,\,\, \Pi_{\gamma}\left(\phi\right)=
\left(\begin{array}{cc} z_{+} & 0 \\ 0 &
\gamma z_{-}
\end{array}\right).
\end{eqnarray*}
The determinants are understood as functional determinants on the
space of two-component functions defined on the product interval
$J\otimes J'$.

Although the formalism in principle can be applied at the spectral
edge, we are primarily interested in the correlation functions in the
bulk of the spectrum. Taking the scaling limit would then greatly
facilitate the calculation. In order to avoid unessential
complications related to the energy dependence of the average density
of states, we replace $\hat{\mathrm{k}}$ with its value
(\ref{scaledk}) in the scaling limit assuming
$\bar{\Delta}\equiv\rho^{-1}\left(\epsilon\right)$ is a constant, and
then regularise $k$ as
\begin{equation}
\label{kreg}
k_{\eta}\left(u-w\right)=\frac{\sin\pi\left(u-w\right)}
{\pi\left(u-w\right)}
e^{-\left(1/2\right)\eta\left(\left|u\right|
+\left|w\right|\right)},
\end{equation}
where the limit $\eta\rightarrow 0$ is implied in all expressions
involving this kernel, and the energy variables
are rescaled by $\bar{\Delta}$.  Using equation (\ref{r}) it is easily
shown that $\langle \mathfrak{n}(J_\infty)\rangle=\langle
\mathfrak{n}(J'_\infty)\rangle=2/\eta$, and
$\langle[\mathfrak{n}(J_\infty)-\mathfrak{n}(J'_\infty)]^2\rangle\sim
O(\eta)$. Thus, although the regularisation formally violates the
level number conservation, the corresponding error tends to zero in
the limit $\eta\rightarrow 0$. At the same time, for any finite value
of $\eta$ the number of levels in the semi-infinite intervals
$J_\epsilon$ and $J'_{\epsilon+\omega}$ is finite, ensuring the
convergence of the integrals over energies at the lower limit.  In the
following we will suppress the index $\eta$.

As written, (\ref{det1}) involves a matrix oscillating integral
kernel defined on a product of semi-infinite intervals. However, as we
will now show for the case $q=0$, it can be rewritten in the form
which is (i) more amenable to numerical analysis, and
(ii) makes the integrability of the kernel (in the sense discussed in
Ref. \cite{Izergin}) manifest. Without loss of generality, we can set 
$\epsilon=0$, and shift the variables so as to define the determinant 
on the quadrant $(-\infty,0]\otimes(-\infty,0]$. The corresponding 
shift operator is absorbed into the redefinition $\hat{\mathcal D}
\rightarrow e^{\omega d/du}\hat{\mathcal D}$. The term 
involving the $\delta$-function in the upper right corner of 
(\ref{kmatrix}) can be separated to reveal the dyadic structure of 
the remainder:
\begin{eqnarray*}
\hat{K}=\left(\begin{array}{c}\mathbb{I}  \\
\hat{\mathcal D}\end{array}\right) \otimes \left( \begin{array}{cc} \hat{k} &
\,\,\,\hat{\mathcal D}^{-1} \hat{k}\end{array}\right)-
\left(\begin{array}{cc}
0 & \hat{\mathcal
D}^{-1}\mathbb{I} \\ 0& 0\end{array}\right),
\end{eqnarray*}
where the we have also used $\hat{\mathcal{D}}^{\dagger}_w
k(u-w)=\hat{\mathcal{D}}_u k(u-w)$.  Now, using the identities
$(\mathbb{I}\sigma_0+z_{-}\hat{\mathcal D}^{-1}
\mathbb{I}\sigma_{+})^{-1}=(\mathbb{I}\sigma_0-z_{-}\hat{\mathcal D}^{-1}
\mathbb{I}\sigma_{+})$, and $\det (\mathbb{I}\sigma_0+z_{-}\hat{\mathcal
D}^{-1}\mathbb{I}\sigma_{+})=1$, where
$\sigma_{+}=(\sigma_1+i\sigma_2)/2$, one obtains
\begin{equation}
\fl \qquad \det\left[\mathbb{I}\sigma_0-\hat{K}\Pi\left(\phi\right)\right]
=\det\left[\mathbb{I}-\left(z_{+}\hat{k}-z_{+}z_{-}\hat{k}
\left(\hat{\mathcal D}^{-1}
\mathbb{I}\right)\hat{\mathcal D}+z_{-}\left(\hat{\mathcal
D}^{-1}\hat{k}\right)\hat{\mathcal D}\right)\right].
\end{equation}
Employing the Fourier representations,
\begin{eqnarray*}
\left\{\begin{array}{c}k(u) \\ \delta(u) \end{array}\right\}
=\frac{1}{2}\int_{-\infty}^{\infty}d\lambda \left\{ \begin{array}{c} 
\theta(1-\lambda)
\theta(1+\lambda) \\ 1\end{array}\right\}  e^{i\lambda\pi u},
\end{eqnarray*}
%
one finds
\begin{eqnarray}
\left[z_{+}\hat{k}-z_{+}z_{-}\hat{k}\left(\hat{\mathcal D}^{-1}
\mathbb{I}\right)\hat{\mathcal D}+z_{-}\left(\hat{\mathcal
D}^{-1}\hat{k}\right)\hat{\mathcal D}\right](u,w)\nonumber \\ 
\fl \qquad \!\! =\int_{-1}^1\!\!\!\frac{d\lambda}{2} e^{i\pi \lambda 
\left(w-u\right)}\left\{z_{+}-\frac{z_{+}z_{-}}{2\pi i}\int_{-\infty}^{\infty} 
\!\!\!d\mu \frac{e^{i\pi w\left(\mu-\lambda\right)}{\mathcal
D}^{-1}\left(\mu\right)}{\lambda-\mu-i\delta}\hat{\mathcal
D}+z_{-}{\mathcal D}^{-1}\left(\lambda\right)\hat{\mathcal D}\right\},
\end{eqnarray}
where in the infinite rank case
\begin{equation}
\mathcal{D}\left(\lambda\right)=\exp\left\{i\pi\omega\lambda+\pi^2
x^2\lambda^2/2\right\},
\end{equation}
and in the finite rank case
\begin{equation}
\mathcal{D}\left(\lambda\right)=\exp\left\{i\pi\omega\lambda\right\}\prod_{\kappa=1}^r\left[1-i
  \lambda \tan\delta_k \right].
\end{equation}
Finally, the cyclic invariance of the determinant and the identity
$z_{+}+z_{-}=z_{+}z_{-}$ are used to perform the integrals in the
$u$-$w$ space, bringing the determinant to the form
\begin{equation}
\label{z0}
Z_0\left(\omega,\mathfrak{x};\phi\right)\equiv
\det\left[\mathbb{I}\sigma_0-\hat{K}\Pi\left(\phi\right)\right]=
\det_{\left[-1,1\right]}\left[\mathbb{I}-\mathcal{K}\left(\phi\right)\right],
\end{equation}
where the matrix elements of the kernel $\mathcal{K}\left(\phi\right)$ are
\begin{equation}
\label{k}
{\mathcal K}\left(\lambda,\mu;\phi\right)=\frac{1}{4\pi i}
\sqrt{{\mathcal D}\left(\lambda\right){\mathcal D}\left(\mu\right)}
\frac{F(\lambda;\phi)-F(\mu;\phi)}{\lambda-\mu},
\end{equation}
and
\begin{equation}
\label{f}
F(\lambda;\phi)=\frac{(z_{+}+z_{-})}{\pi i}
\,\dashint_{-\infty}^{\infty}\!\! d\mu\frac{{\mathcal
D}^{-1}(\mu)}{\mu-\lambda}-(z_{+}-z_{-}) \,{\mathcal D}^{-1}(\lambda)
\end{equation}
Here the integral is understood in the sense of the Cauchy principal
value, with variables $\lambda$ and $\mu$ restricted to the interval
$\left[-1,1\right]$. We employ the notation $\mathfrak{x}$ to denote the
parametric dependence of $Z$ in both finite and infinite rank
cases. A similar, although somewhat more cumbersome, expression, is
easily reconstructed for the contribution of
$\det\left[\mathbb{I}\sigma_0-\hat{K}\Pi_\gamma\left(\phi\right)\right]$
in the $q\ne 0$ case.

\begin{figure}[h]
\begin{center}
\includegraphics[width=3.4in]{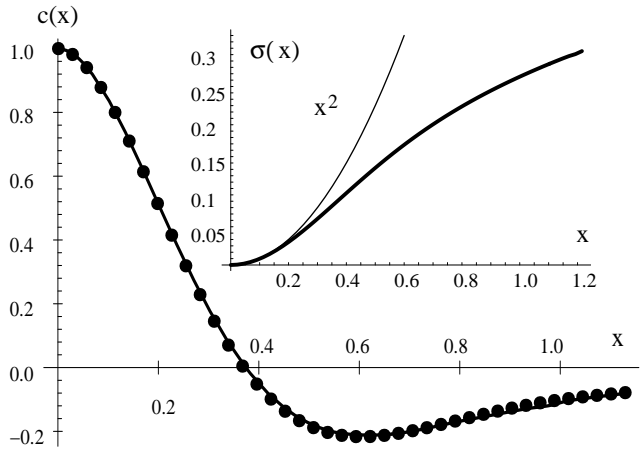}
\caption[]{\label{fig1}Single level universal velocity autocorrelation
function obtained from (\ref{cq}), (\ref{z0}), (\ref{k}), and
(\ref{f}) (solid line) vs. direct numerical simulation of large random
matrices~\cite{eduardo} (dots). The width $\sigma\left(x\right)$ of
the Gaussian distribution of level shifts together with the $x^2$
asymptotics at small $x$ is shown in the inset.}
\end{center}
\end{figure}

Equations (\ref{z0}), (\ref{k}), and (\ref{f}) form the central
results of this section. In the remainder we will consider several
applications of these equations to the calculation of number-dependent
correlation functions. The integral of the generating function itself
$\int d\phi Z_0\left(\omega,\mathfrak{x};\phi\right)/2\pi$ gives
the probability $\mathfrak{P}_0(\omega,\mathfrak{x})\equiv
\mathfrak{P}_0\left(J_\epsilon,J_{\epsilon+\omega}\right)$. In order
to obtain the correlation functions $P_q$ and $C_q$ introduced above, we
rewrite their definitions in the form
\begin{eqnarray}
P_q\left(\omega,\mathfrak{x}\right)&&=\left\langle
\partial_\epsilon\mathfrak{n}\left(J_\epsilon\right)
\partial_\epsilon\mathfrak{n}\left(J'_{\epsilon+\omega}\right)
\delta_{\mathfrak{n}\left(J_\epsilon\right),\mathfrak{n}\left(J'_{\epsilon+\omega}\right)+q}\right\rangle,
\\
C_q\left(\mathfrak{x}\right)&&=\left\langle \partial_\mathfrak{x}\mathfrak{n}\left(J_\epsilon\right)
\partial_\mathfrak{x}\mathfrak{n}\left(J'_{\epsilon+\omega}\right)
\delta_{\mathfrak{n}\left(J_\epsilon\right),\mathfrak{n}\left(J'_{\epsilon+\omega}\right)+q}\right\rangle.
\end{eqnarray}
Using translational invariance both in the energy and the parameter
spaces, rescaling the energies by $\bar{\Delta}$, and rescaling the
variables in $\mathfrak{x}$ by $C_0^{-1/2}$ \cite{SA}, we obtain for
the rescaled correlation functions $p_q$ and $c_q$ \cite{SS}
\begin{eqnarray}
\label{finalpq}
p_q\left(\omega,\mathfrak{x}\right)&&=\left(-1\right)^q
\int_0^{2\pi}\frac{d\phi}{2\pi}
\frac{e^{iq\phi}}{z_+z_-}\left(-\partial^2_\omega\right)
Z_q\left(\omega,\mathfrak{x};\phi\right), \\
\label{cq}
c_q\left(\mathfrak{x}\right)&&=\left(-1\right)^q
\int_{-\infty}^{\infty}d\omega \int_0^{2\pi}\frac{d\phi}{2\pi}
\frac{e^{iq\phi}}{z_+z_-} \left(-\partial_\mathfrak{x}\right)^2
Z_q\left(\omega,\mathfrak{x};\phi\right).
\end{eqnarray}
Let us now consider several special cases. (i) In the infinite rank
case $\mathfrak{x}$ reduces to a single parameter
$x=\left(\mathrm{Tr}\, \hat{\mathrm{v}}^2\right)^{1/2}/\pi$.  The
single level universal velocity autocorrelation function
$c\left(x\right)\equiv c_0\left(x\right)$ is plotted in
Fig. \ref{fig1}. For comparison we also show the results of direct
numerical simulation of large random matrices \cite{eduardo}. We also
find, based on numerical evaluation of equations (\ref{finalpq}),
(\ref{z0}), (\ref{k}), and (\ref{f}) that the function
$p\left(\omega,x\right)\equiv p_0\left(\omega,x\right)$ -- the
distribution of parametric single level shifts -- has a Gaussian form
$p\left(\omega,x\right)=\exp\left\{-\omega^2/
2\sigma\left(x\right)\right\}/\sqrt{2\pi\sigma\left(x\right)}$, in
accordance with the earlier conjecture \cite{vallejos}. An analytical
proof of this statement, as well as finding an analytical expression
for the width $\sigma\left(x\right)$ (shown in the inset in
Fig. \ref{fig1}) are some of the open questions posed by the results
of the present investigation.

(ii) In the finite rank case further progress can be made
due to a simple pole structure of
$\mathcal{D}^{-1}\left(\lambda\right)$. Leaving a detailed
consideration of the finite rank case to a future publication, we
present here the results for $c_0$ and $p_0$ in the $r=1$
case. $\mathfrak{x}$ is now parametrised by a single phase shift
$\delta$. In Ref. \cite{SMS} the distribution $p_0$ for $r=1$ was
found using the fact that, as a special property of $r=1$, this
distribution coincides with the level spacing of the {\em combined}
ensemble of levels
$\left\{\epsilon_\alpha\right\}\cup\{\epsilon'_\beta\}$.
Using the present formalism, this quantity, as well as the
corresponding $c_0$ can be evaluated in a more direct way. Introducing
an auxiliary {\em non-parametric} kernel
\begin{equation}
\mathfrak{k}\left(\lambda,\mu\right)=\frac{1}{2\pi i}\frac{1- 
\exp\left\{i\pi\omega\left(\lambda-\mu\right)\right\}}{\lambda-\mu},
\end{equation}
its Green function
$g\left(\lambda,\mu\right)=\left(\mathbb{I}-\mathfrak{k}\right)^{-1}$, and
the determinant
$d\left(\omega\right)=\det\left(\mathbb{I}-\mathfrak{k}\right)$, we
find (assuming $\delta>0$)
\begin{equation}
\fl c^{\left(r=1\right)}_0\left(\delta\right)=\frac{d^2}{d\delta^2}\int_0^{\infty}\frac{d\omega}{2\pi}d\left(\omega\right)
\exp\left\{-\pi\omega\arctan\delta\right\}\int_{-1}^{1}d\lambda d\mu 
\frac{g\left(\lambda,\mu\right)\exp\left\{i\pi\omega\mu\right\}}{\arctan\delta-i\lambda}.
\end{equation}
A somewhat more lengthy expression is obtained for $p_0$:
\begin{eqnarray}
\fl  p^{\left(r=1\right)}_0\left(\omega;\delta\right)=\theta\left(\omega\right)\frac{\pi
  d\left(\omega\right)}{2}\exp\left\{-\pi\omega\arctan\delta\right\}
  \int_{-1}^{1}d\lambda d\mu
  \frac{g\left(\lambda,\mu\right)\exp\left\{i\pi\omega\mu\right\}}{1-i\lambda\tan\delta}
  \nonumber \\
  \times \left\{\frac{\left(1-i\mu\tan\delta\right)^2}{\tan\delta}
  +\left(1-i\mu\tan\delta\right)t\left(\omega\right)-T\left(\omega,\delta\right)\right\},
\end{eqnarray}
where $t\left(\omega\right)=\int_{-1}^{1}d\lambda d\mu
g\left(\lambda,\mu\right)\exp\left\{i\pi\omega\left(\mu-\lambda\right)\right\}$, 
and 
\begin{equation}
T\left(\omega,\delta\right)=\int_{-1}^{1}d\lambda d\mu
g\left(\lambda,\mu\right)\exp\left\{i\pi\omega\left(\mu
-\lambda\right)\right\}\left(1-i\mu\tan\delta\right). \nonumber
\end{equation}
Up to a simple transformation, the kernel $\mathfrak{k}$ is identical
to the kernel which controls the {\em non-parametric} level
spacing distribution $\mathfrak{S}_1$. In light of the ``interleaved
combs'' constraint on the evolution of levels when $r=1$, such a
connection to $\mathfrak{S}_1$ is not unexpected.

Within the framework presented here various number-dependent
correlation functions in the unitary ensembles can be computed ``at
will''. Nevertheless, several open questions remain unanswered, and
their resolution would greatly enhance our understanding of parametric
evolution of random matrices. As already mentioned, the Gaussian form
of $p_0$ in the $r\to\infty$ case is probably an indicator that the
$\phi$ integration can be performed analytically in some form in the
infinite rank case as well. If so, a more ``processed'' expression for
$c_0^{r\to\infty}$ may emerge, throwing more light on the properties
of the kernel $\mathcal{K}\left(\phi\right)$. Separately, an analysis
of the asymptotic behaviour of the generating functions along the lines
of Ref. \cite{tracy} is a natural way to develop the
present formalism further. Finally, the technique of ``counting'' the
number of levels between the spectral edge and a point in the bulk of
the spectrum by means of introducing a regularised kernel should be
applicable to ensembles with different symmetries, so that if an
analogue of (\ref{corrdet}) is obtained for such ensembles, it is
likely to open the door to the computation of the corresponding
number-dependent correlation functions.

\end{document}